\documentclass[twocolumn]{aastex61}

\newcommand\aastex{AAS\TeX}

\usepackage{mathtools}

\submitjournal{ApJ}

\shorttitle{\aastex\ Revisting The Extended Schmidt Law}
\shortauthors{Shi et al.}

\begin{document}

\title{Re-visiting The Extended Schmidt Law: the important role of existing stars in regulating star formation}

\correspondingauthor{Yong Shi}
\email{yshipku@gmail.com}

\author[0000-0002-8614-6275]{Yong Shi}
\affil{School of Astronomy and Space Science, Nanjing University, Nanjing 210093, China.}
\affil{Key Laboratory of Modern Astronomy and Astrophysics (Nanjing University), Ministry of Education, Nanjing 210093, China.}

\author{Lin Yan}
\affil{Infrared Processing and Analysis Center, California Institute of Technology, 1200 E. California Blvd, Pasadena, CA 91125, USA}

\author{Lee Armus}
\affil{Infrared Processing and Analysis Center, California Institute of Technology, 1200 E. California Blvd, Pasadena, CA 91125, USA}

\author{Qiusheng Gu}
\affil{School of Astronomy and Space Science, Nanjing University, Nanjing 210093, China.}
\affil{Key Laboratory of Modern Astronomy and Astrophysics (Nanjing University), Ministry of Education, Nanjing 210093, China.}

\author{George Helou}
\affil{Infrared Processing and Analysis Center, California Institute of Technology, 1200 E. California Blvd, Pasadena, CA 91125, USA}

\author{Keping Qiu}
\affil{School of Astronomy and Space Science, Nanjing University, Nanjing 210093, China.}
\affil{Key Laboratory of Modern Astronomy and Astrophysics (Nanjing University), Ministry of Education, Nanjing 210093, China.}

\author{Stephen Gwyn}
\affil{NRC-Herzberg Astronomy and Astrophysics, National Research Council of Canada, 5071 West Saanich Road, Victoria, BC V9E 2E7, Canada}

\author{Sabrina Stierwalt}
\affil{Department of Astronomy, University of Virginia, P.O. Box 400325, Charlottesville, VA 22904, USA}

\author{Min Fang}
\affil{Department Of Astronomy And Steward Observatory, University of Arizona, 933 N Cherry Ave, Tucson, AZ 85721, USA}

\author{Yanmei Chen}
\affil{School of Astronomy and Space Science, Nanjing University, Nanjing 210093, China.}
\affil{Key Laboratory of Modern Astronomy and Astrophysics (Nanjing University), Ministry of Education, Nanjing 210093, China.}

\author{Luwenjia Zhou}
\affil{School of Astronomy and Space Science, Nanjing University, Nanjing 210093, China.}
\affil{Key Laboratory of Modern Astronomy and Astrophysics (Nanjing University), Ministry of Education, Nanjing 210093, China.}

\author{Jingwen Wu}
\affil{National Astronomical Observatories, Chinese Academy of Sciences, 20A Datun Road, Chaoyang District, Beijing, China}

\author{Xianzhong Zheng}
\affil{Purple Mountain Observatory, Chinese Academy of Sciences, 2 West Beijing Road, Nanjing 210008, China}

\author{Zhi-Yu Zhang}
\affil{Institute for Astronomy, University of Edinburgh, Royal Observatory, Blackford Hill, Edinburgh EH9 3HJ, UK}

\author{Yu Gao}
\affil{Purple Mountain Observatory, Chinese Academy of Sciences, 2 West Beijing Road, Nanjing 210008, China}

\author{Junzhi Wang}
\affil{Shanghai Astronomical Observatory, Chinese Academy of Sciences, 80 Nandan Road, Shanghai 200030, China}




\begin{abstract}

  We revisit the proposed extended Schmidt law (Shi et al. 2011) which
  points that the star formation efficiency in galaxies depends on the
  stellar mass surface density,     by      investigating
  spatially-resolved  star  formation  rates (SFRs),  gas  masses  and
  stellar masses of star formation regions in a vast range of galactic
  environments, from the outer disks of dwarf galaxies to spiral disks
  and to  merging galaxies as  well as individual molecular  clouds in
  M33.   We  find  that  these  regions are  distributed  in  a  tight
  power-law       as      $\Sigma_{\rm       SFR}{\propto} ({\Sigma_{\rm
      star}^{0.5}}{\Sigma_{\rm  gas}})^{1.09}$, which  is also  valid for  the
  integrated measurements  of disk  and merging galaxies  at high-$z$.
  Interestingly,  we show  that star  formation regions  in the  outer
  disks of  dwarf galaxies with  $\Sigma_{\rm SFR}$ down  to 10$^{-5}$
  M$_{\odot}$/yr/kpc$^{2}$,    which    are     outliers    of    both
  Kennicutt-Schmidt  and  Silk-Elmegreen   law,  also follow  the  extended
  Schmidt law.  Other  outliers in the Kennicutt-Schmidt  law, such as
  extremely-metal poor star-formation regions, also show significantly
  reduced  deviations from  the  extended Schmidt  law.  These  results
  suggest  an important  role for  existing stars  in helping to regulate  star
  formation through the  effect of  their gravity  on the  mid-plane
  pressure in a wide range of galactic environments.

\end{abstract}





\section{Introduction} \label{sec:intro}

Stars  are born  in clouds of cold gas. Understanding
how  gas is  converted into  stars is  crucial for  constraining theories of galaxy
formation and evolution across cosmic time. In the early universe,
primordial gas  with little or no  metals collapse into the  first and
second generation  stars whose  radiation is responsible for re-ionizing the Universe.
The subsequent rise in the  cosmic star formation rate (SFR) density
until $z\sim$2 followed by the  decline to $z$=0
has   been  well mapped in a broad sense by modern, multi-wavelength surveys
\citep[e.g.][]{Madau14}.   The driving  mechanism of  this cosmic  SFR
evolution is thought to be related  to the cold gas reservoir available
within and surrounding   galaxies   and the physical conditions which
allow this gas to be converted into new stars \citep{Genzel10, Tacconi13}.

Empirical relationships between  the star formation rate  and cold gas
mass (or  star formation law)  as first proposed  by \citet{Schmidt59}
are powerful  tools for understanding star  formation and  also useful
prescriptions  for  creating  stars  from   cold  gas  in   cosmological
simulations. Since the pioneering  work of \cite{Kennicutt89},
studies have established the power-law
form of the relationship between the SFR surface density ($\Sigma_{\rm gas}$)
and gas (atomic+molecular) mass surface density ($\Sigma_{\rm
  gas}$), i.e.,  the Kennicutt-Schmidt law  in a form  of $\Sigma_{\rm
  SFR}$ $\propto$ $\Sigma_{\rm  gas}^{N}$ with $N$ around  1.4.  It is
found that  for star formation  regions in individual galaxies
the relationship show variations in both slope and normalization
\citep{Kennicutt07, Leroy08,  Shi11, Boquien11},  but for 
ensembles of  galaxies covering a  large dynamic range of  SFRs, the
relationship has a relatively small dispersion of about 0.3  dex
\citep{Kennicutt98, Wyder09,  Shi11}.  This  suggests that the  gas mass
surface density  is the primary  factor in regulating the  SFR surface
density.

Deviations from the Kennicutt-Schmidt law occur at the low density end
($\Sigma_{\rm gas}$ $\lesssim$ 10 M$_{\odot}$/pc$^{2}$ or $\Sigma_{\rm
  SFR}$  $\lesssim$  10$^{-3}$  M$_{\odot}$/yr/kpc$^{2}$)  where  star
formation is systematically below the power-law extrapolation from the
higher gas  density regime, including  low-surface-brightness galaxies
and the  outer disks of galaxies  \citep{Kennicutt98, Bigiel08}.  Such
deviation implies an additional parameter beside the gas mass that may
help  to  regulate ongoing  star  formation.   At least  two  modified
versions of the  Schmidt relationship have been proposed.   One is the
so-called  Silk-Elmegreen  relationship  that  invokes  the  dynamical
orbital  timescale   as  $\Sigma_{\rm  SFR}$   $\propto$  $\Sigma_{\rm
  gas}$/$t_{\rm  dyn}$ \citep{Silk97,  Elmegreen97}.   Another is  the
extended Schmidt law that invokes  the stellar mass surface density as
$\Sigma_{\rm    SFR}$   $\propto$    ${\Sigma_{\rm   gas}}{\Sigma_{\rm
    star}^{N}}$,  with   $N$  around   0.5  \citep{Shi11}.    In  both
relationships,  the low  surface brightness  (LSB) galaxies  and outer
disks of  spiral galaxies  along with  the main  disks of  spirals and
star-burst galaxies define a power  law across the whole dynamic range
with no  obvious break  at the  low density  end down  to $\Sigma_{\rm
  SFR}$ $\sim$ 10$^{-4}$ M$_{\odot}$/yr/kpc$^{2}$.   In this study, we
will include measurements of star formation regions in the outer disks
of dwarf galaxies with $\Sigma_{\rm SFR}$ ten times lower than LSB galaxies and
outer disks of  spirals to further test these  two alternative models,
and thus to establish which parameter (dynamical timescale vs. stellar
mass) in  addition to the  gas mass  plays a more  important secondary
role in regulating star formation.

Besides  LSB galaxies  and  galaxy outer  disks,  deviations from  the
Kennicutt-Schmidt law are  also seen for other types  of galaxies. For
example,  individual star  formation regions  in extremely  metal-poor
galaxies with metallicities below 10\% of Solar show significant lower
SFRs than those in spiral galaxies  at given gas mass densities in the
Kennicutt-Schmidt plane \citep[e.g.][]{Shi14}.  The disk-averaged SFRs
of   low   mass   metal-poor    galaxies   also   deviate   from   the
Kennicutt-Schmidt law  \citep[e.g.][]{Roychowdhury15}.  In this study,  we will examine
if these  galaxies show reduced  deviations from the  extended Schmidt
law once including the stellar mass term.

With the spatially-resolved observations of nearby galaxies as enabled
by multi-wavelength modern  facilities, we can also  evaluate the star
formation law on sub-kpc scales  in different galactic environments.
Such  measurements relate  the local  SFRs to  {\it in  situ} gas  and
stellar mass that may be physically related to current star formation.
The paper is  organized as following. In  \S~\ref{sec:data} we present
the  data and  measurements of  physical parameters  including stellar
mass, gas  mass and SFR surface  density.  The updated version  of the
extended   Schmidt   law   is  presented   in   \S~\ref{sec:ES}.    In
\S~\ref{sec:outlier}, we  discuss the distribution of  the outliers of
the Kennicutt-Schmidt law in the extended Schmidt plane. The origin of
the  extended Schmidt  law  is discussed  in \S~\ref{sec:model}.   The
conclusion is present in \S~\ref{sec:conclusion}.

\section{Data and Measurements of Physical Parameters} \label{sec:data}

\subsection{Outer Disks Of Three Dwarf Galaxies} 


To  test the  role of  the stellar  mass in  regions of  extremely low
$\Sigma_{\rm SFR}$, we  selected three dwarf galaxies (Ho II,  Ho I \&
DDO 154) that have  deep HI data as observed in the  program of the HI
Nearby Galaxy  Sample \citep{Walter08} as  well as deep  far-UV images
from the  GALEX data archive.   The HI images were  naturally weighted
with a  beam size  of $\sim$  14$\arcsec$x12.5$\arcsec$ and  a typical
column  density sensitivity  of 9$\times$10$^{19}$  cm$^{-2}$ at  this
resolution.   For  these  three  galaxies,  single  dish  measurements
indicate  that  there  is  no  significant missing  HI  flux  at  this
resolution \citep{Walter08}.  The GALEX image has a spatial resolution
about 5$''$  \citep{Morrissey07}.  The archive-produced sky  image was
subtracted from the  target image, with the  1-$\sigma$ scatter around
(1-2)$\times$10$^{-8}$   Jy  at   the  pixel   scale  of   1$\farcs$5,
corresponding   to   a   SFR  surface   density   of   (1-2)x10$^{-5}$
M$_{\odot}$/yr/kpc$^{2}$.

The  optical  observations of  the  above  three dwarf  galaxies  were
carried out at $g$ and $r$ band  with the MegaCAM of CFHT (PI: Y. Shi)
through  China Telescope  Access Program.   The observing  details are
listed in  Table~\ref{tab_cfht} with a  typical seeing of  0$\farcs$8 in
the  $g$-band and  0$\farcs$6 in  the $r$-band.   To best  recover the
faint  outer disk  emission  of our  galaxies,  the observations  were
dithered  using the  LDP-CCD-7 mode  as developed  by the  CFHT Elixir
team.  The dither uses large offsets ($\sim$ 15$'$) to move the target
across the MegaCAM  mosaic such that the underlying  background of the
target out to the dither size can be recovered and subtracted from the
series of exposures.  The outer disks of our three galaxies extend out
to 5-10$'$, smaller than the dither size so that the full advantage of
the above  dither pattern  can be  fulfilled.  Individual  frames were
first reduced  with the Elixier  pipeline \citep{Magnier04}  including the
bias  subtraction,   the  flat  field  response   correction  and  the
application  of the  relative flux  calibration.  Then  the individual
frames  were  mosaiced by  taking  care  of  diffuse light  using  the
MegaPipe image  processing pipeline  \citep{Gwyn08} that  includes the
image  grouping,  astrometric  calibration,  photometric  calibration,
image stacking and catalog generation.

\begin{figure*}
  \centerline{\includegraphics[width=1.0\textwidth]{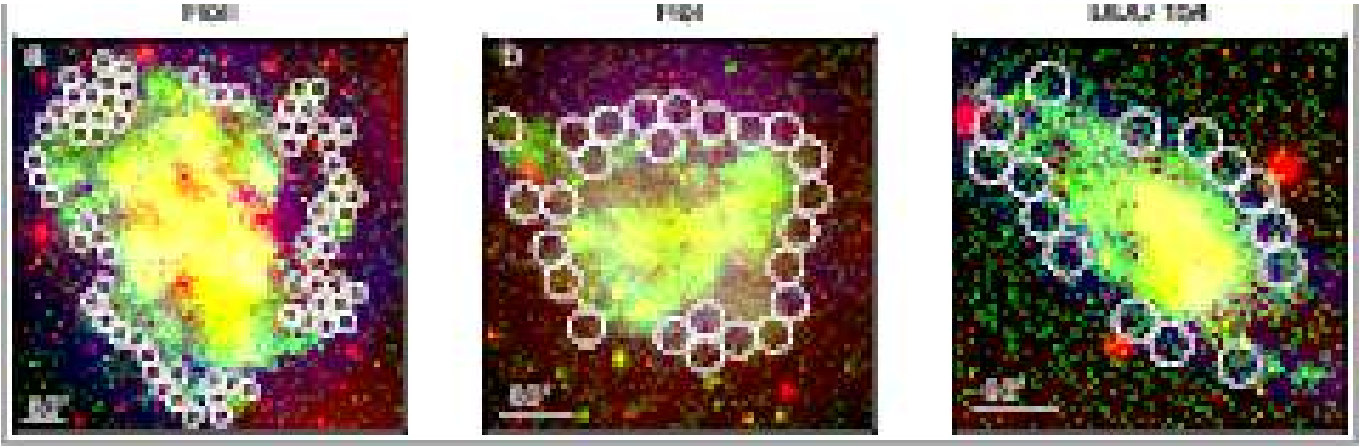} }
\caption{\label{fig_dwarf_od} The color images of three dwarf galaxies composed of
  the $g$-band (red), the far-UV (green) and  the HI atomic gas  (blue) images. Individual
white circles indicate star-forming regions in the outer disks with low SFR
surface density ($<$ 10$^{-4}$ M$_{\odot}$/yr/kpc$^{2}$). }
\end{figure*}

To probe star-forming regions  with $\Sigma_{\rm SFR}$ below 10$^{-4}$
M$_{\odot}$/yr/kpc$^{2}$  where the previous  study has  not yet  probed
\citep{Shi11},  we  defined  the  outer disks  as  between  two
contours  at the  SFR  surface densities  of  10$^{-4}$ and  10$^{-5}$
M$_{\odot}$/yr/kpc$^{2}$, respectively. We then  filled the outer disk
with individual  circle regions  of 30$''$ as  outer-disk star-forming
regions as shown  in Fig.~\ref{fig_dwarf_od}.  In total 74,  25 and 15
regions are identified for Ho II, Ho I and DDO 154, respectively.

The surface  densities of SFR, gas mass and
stellar  mass are measured with  the following  procedure.  As  stated
above, the archived  produced background image is  subtracted from the
target  image,  based  on  which  the  far-UV  surface  brightness  is
converted to the SFR surface density with a Kroupa initial
mass function  (IMF) using the following formula \citep{Leroy08}:
\begin{equation}
   \Sigma_{\rm SFR} = 8.1\times10^{-2}{\rm cos}(i)I_{\rm FUV}
\end{equation}
where $\Sigma_{\rm SFR}$ is in M$_{\odot}$/yr/kpc$^{2}$, $I_{\rm FUV}$
is in MJy/sr and $i$ is the  inclination angle of the galaxy  measured at $R_{25}$.
 Note that  the  HI  warp is  seen  in DDO  154  but  all of  our
outer-disk star-forming regions are within the radius where the warp
occurs \citep{Carignan89}.  No  warps are seen in  the remaining two
galaxies (HoI and HoII) \citep{Holwerda11}. As a result, the
adopted inclination at $R_{25}$ is a reasonable value to use for the HI surface
density estimate. No
dust  extinction  is  assumed  for   these  outer  disk  regions.  The
associated  uncertainty  is  the   quadratic  sum  of  the  photometric
uncertainty plus the systematic uncertainty of 0.3 dex.

The total gas  mass of each star-forming region is the HI gas mass multiplied
with a factor  of 1.36 to account for Helium following the formula \citep{Walter08}:
\begin{equation}\label{eqn_hi_gas}
\Sigma_{\rm atomic{\textendash}gas}
=1.20\times10^{4}{\rm cos}(i)(1+z)^{3}\frac{\rm arcsec^{2}}{\rm bmaj{\times}bmin}S_{\rm HI}{\Delta}v
\end{equation}
where  $i$  is  the  inclination angle  of  the  galaxy, $z$ is the redshift of the galaxy,
$\Sigma_{\rm atomic{\textendash}gas}$    is   in M$_{\odot}$/pc$^{2}$,   $S_{\rm HI}{\Delta}v$ is in Jy$\cdot$km/s/beam,
$b_{\rm maj}$ and $b_{\rm min}$ is the major and
minor  beam  sizes  in  arcsec,  respectively.   Here  we  assumed  no
molecular gas in the outer-disk regions. The error is assumed to be 0.2 dex
to account for the flux calibration uncertainty.

To derive  the stellar mass,  we first  measured the $g$  and $r$-band
photometry by summing the fluxes of individual detected sources within
each 30$''$ aperture based on the catalog produced by the MegaPipe.  Here the
high  spatial resolution  (0$\farcs$6-0$\farcs$8)  combined with  deep
exposures enable  to resolve  individual sources within  the aperture of each region.
For one  field (DDO 154) covered  by the Sloan Digital  Sky Survey, we
compared  the MegaPipe  photometry of  all  point sources  in the  one
square  degree  field  to  those  by  SDSS  and  found  no  systematic
difference  ($<$  0.02  mag).  We  also performed   aperture
  photometry by summing  all photons within the aperture of each region in the image
  after  subtracting  the  global   sky  background,  and  a  similar
  measurement  but  after   subtracting both  global  and  local
  backgrounds. These two methods give  overall similar  distributions of
  the stellar masses as compared to the first method.  This indicates that
the light  within the aperture  is dominated  by the point  sources so
that all methods  give similar results.   We adopted  the results
using the first  method that is not subject to  the uncertainty caused
by the  global sky background subtraction  over a large field  of view
and thus has a higher photometric accuracy.

To obtain the stellar mass from the $g$ and $r$ photometry, we  used the  Sloan MPA/JHU
catalog \citep{Brinchmann04}   to   obtain   the   dependence   of   the
mass-to-light ratio at $r$-band on the $g$-$r$ color:
\begin{equation}
  {\rm log}M_{\rm star}+M_{\rm r-band}/2.5 = 1.13+1.49(g-r),
\end{equation}
where  $M_{\rm star}$  is  the stellar  mass  in M$_{\odot}$,  $M_{\rm
  r-band}$  is the  $r$-band absolute  AB magnitude.   As the  $g$-$r$
color of the  majority of the SDSS galaxies lies  between 0.1 and 0.9,
for  outer-disk  regions   with  $g$-$r$  $>$  0.9, we fixed the mass-to-light
ratio to that for a population with $g$-$r$=0.9.   Similarly   the
mass-to-light ratio of regions with $g$-$r$  $<$ 0.1 was fixed to that
for a population with $g$-$r$=0.1. For about 60\% of the regions with $g$-$r$ between 0.1
and 0.9,  the mass-to-light ratio  has a scatter  of 0.25 dex  and the
remaining 40\% outside  the range is assumed to have  a scatter of 0.5
dex. The final error of the  stellar mass, mainly between 0.25 and 0.5
dex, includes the photometric error in the $g$ and $r$ band as well as
the error  in the  mass-to-light ratio.  Our  SDSS-based mass-to-light
ratio  in  the $r$-band  as  a  function  of  $g$-$r$ color  is  quite
consistent with the result derived  from the 50,000 stellar population
synthesis models  \citep{Zibetti09}, with  a difference less  than 0.1
dex for $g$-$r$ between 0.1 and 0.9.

The orbital  dynamical time of Ho  II and DDO  154 is based on  the HI
rotation curve available in \citet{Oh15}.

\subsection{Local Spiral Galaxies}

The spatially-resolved  measurements of  SFRs, gas masses  and stellar
masses of 12  spiral galaxies are carried out  basically following the
literature work \citep{Shi11, Bigiel08}.  The  aperture size is set to
be  750$\times$750  pc$^{2}$  that  can be  achieved  by  the  angular
resolution of  the data at  the largest  distance of the  sample.  The
SFRs are the combinations of the far-UV from the GALEX archive and the
infrared    data   from    The    SIRTF    Nearby   Galaxies    Survey
\citep{Kennicutt03}, based on the equation \citep{Leroy08}:
\begin{equation}\label{eqn_24um_fuv_sfr}
\Sigma_{\rm SFR} = 8.1\times10^{-2}{\rm cos}(i)I_{\rm FUV} + 3.2\times10^{-3}{\rm cos}(i)I_{24{\mu}m},
\end{equation}
where $i$ is the inclination angle of the galaxy, $\Sigma_{\rm SFR}$  is  in  M$_{\odot}$/yr/kpc$^{2}$, and  both
$I_{\rm FUV}$ and $I_{24{\mu}m}$ are in  MJy/sr.  The HI data are from
the H  I Nearby Galaxy  Survey \citep{Walter08} and CO data  are carried
out by  the BIMA SONG  CO J =  1-0 map \citep{Helfer03}, from  which the
total   gas   is   derived as  $\Sigma_{\rm   gas}$   =   $\Sigma_{\rm atomic{\textendash}gas}$+$\Sigma_{\rm    mol{\textendash}gas}$.    The
atomic gas mass follows the equation  (2). The molecular part is based
on
\begin{eqnarray}\label{eqn_co_gas}
 \Sigma_{\rm mol{\textendash}gas} =
 \nonumber 393{\rm cos}(i)\frac{\rm arcsec^{2}}{\rm bmaj{\times}bmin}\frac{X_{\rm CO}}{(2\times10^{20} \frac{\rm cm^{-2}}{\rm K{\cdot}km/s})}\\ 
 (1+z)^{3}S_{\rm CO}{\Delta}v,
\end{eqnarray}
where $i$ is the inclination angle of the galaxy, $\Sigma_{\rm mol{\textendash}gas}$  is in  M$_{\odot}$/pc$^{2}$
and $S_{\rm  CO}{\Delta}v$ is in Jy${\cdot}$km/s/beam.   Here a factor
of 1.36 has  been included to account for the  Helium. The CO emission
is the $J$=1-0 and  we adopt $X_{\rm CO}$=2$\times$10$^{20}$ cm$^{-2}$
for spiral galaxies.

Based on the available  Spitzer 3.6 $\mu$m data \citep{Kennicutt03}, the stellar
mass is derived by the formula \citep{Leroy08}:
\begin{equation}\label{eqn_irac_mstar}
   \Sigma_{\rm star}=140{\rm cos}(i)I_{\rm 2mass-K}=280{\rm cos}(i)I_{3.6{\mu}m}
\end{equation}
where $i$ is the inclination angle of the galaxy, $\Sigma_{\rm star}$ is in M$_{\odot}$/pc$^{2}$ and both $I_{\rm 2mass-K}$
and $I_{3.6{\mu}m}$ is in MJy/sr. For five galaxies that have SDSS optical images, we also derived
the stellar mass using Equation 3 and found the difference from the IRAC-based one is smaller than 0.2 dex.

\subsection{Luminous Infrared Galaxies In The Local Universe}

{\noindent \bf NGC 1614}


NGC  1614  is a  LIRG  with  an  infrared luminosity  of  10$^{11.65}$
$L_{\odot}$ \citep{Armus09}  at the late  stage of a  merging process.
It likely has experienced a  minor merging event that triggers intense
star  formation  in  a  circumnuclear ring  region.   We  carried  out
spatially-resolved  measurements for  the  nuclear region  as well  as
individual regions in the star-forming  ring of NGC 1614. The aperture
size  is  about  100-200  pc  that is  about  two  times  the  angular
resolution.

The total gas mass is measured based  on the CO 6-5 map as observed by
ALMA  \citep{Xu15}   with  a  synthesized  beam   of  0.26$\times$0.20
arcsec$^{2}$.   We  retrieved  the  data from  the  ALMA  archive  and
obtained the  intensity map by  integrating the velocity from  4500 to
5000 km/s.   To estimate the molecular  gas mass from CO  6-5, we used
the formula  in the  literature \citep{Xu15}  with a  slightly smaller
$X_{\rm  CO}$: CO  6-5 is  first  converted to  CO 2-1  by assuming  a
brightness temperature  ratio of 0.72  between the two; to  derive the
factor from  CO 2-1  to the  molecular gas mass,  we converted  the CO
(2-1) flux  \citep{Konig13} to CO  (1-0) by  dividing a factor  of 3.1 (0.78 in the brightness temperature)
\citep{Wilson08} and  then used the  conversion factor $X_{\rm  CO}$ =
2$\times$10$^{20}$ cm$^{-2}$ for  CO 1-0 with an  additional factor of
1.36 for the helium. The final formula is the following:
\begin{equation}
 \Sigma_{\rm gas} = 13.5{\times}f_{\rm CO(6-5)},
\end{equation}
where $\Sigma_{\rm gas}$ is in M$_{\odot}$/pc$^{2}$ and $f_{\rm CO(6-5)}$ is in
Jy$\cdot$km/s/arcsec$^{2}$.

SFRs are measured based the  radio continuum map. We retrieved the
radio 8.4 GHz  (X-band) and 4.86 GHz (C-band) intensity  maps from the
VLA archive that  have synthesized beam sizes  of 0.26$\times$0.21 and
0.60$\times$0.40 arcsec$^{2}$,  respectively.  With a Kroupa  IMF, the
SFR is based on the formula \citep{Murphy12}:
\begin{eqnarray}
 {\rm SFR} =
 \frac{10^{-27}}{2.18(\frac{T_{\rm e}}{10^{4} K})^{0.45}(\frac{\nu}{\rm GHz})^{-0.1}+15.1(\frac{\nu}{\rm GHz})^{-\alpha^{\rm NT}}}L_{\nu}
\end{eqnarray}
where SFR  is in M$_{\odot}$/yr,  $T_{\rm e}$ is the  thermal electron
temperature at 10$^{4}$ K, $\alpha^{\rm  NT}$ is the non-thermal radio
spectral index  with a suggested  value of  0.85, and $L_{\nu}$  is in
erg/s/Hz.  With the $\alpha^{\rm NT}$=0.85,  the two SFRs based on the
X-band and  C-band radio  data are quite  consistent with  each other,
with a median offset  of 0.12 dex. The final result  is the mean value
of the two SFRs. Note that because of different formula as used in the
literature work \citep{Xu15}, our SFRs are about a factor of two lower.

To measure the  stellar mass, we retrieved the archived  images of the
{\it Hubble} Space  Telescope in the filters  of ACS-F435W, ACS-F814W,
NICMOS-F110W, NICMOS-F160W and NICMOS-F222M, whose spatial resolutions
are comparable  to those  of CO  6-5 and  radio continuum  images. The
astrometry of each NICMOS image is  first registered to that of ACS by
matching  several  field  stars.   Then all  HST  images  are  further
registered by  matching the  optical center to  the radio  center. The
broad-band SED within each aperture  is then constructed and fitted to
derive the stellar synthesis model \citep{Brinchmann04}.\\

{\bf IC 4687}

IC 4687 is  a LIRG with an IR luminosity  about 10$^{11.3}$ $L_{\odot}$
\citep{Pereira-Santaella11}. The spatially resolved measurements
of  SFRs and  gas masses  of  IC 4687  have  been carried  out in  the
literature study \citep{Pereira-Santaella16}.  The SFRs are based on the
extinction-corrected Pa$\alpha$  emission as observed by  HST, and the
gas  mass is  based on  CO  J=2-1 emission  as observed  by ALMA.  The
conversion factor is  taken to be $X_{\rm  CO}$ =2x10$^{20}$ cm$^{-2}$
as   argued   in  the   literature   study \citep{Pereira-Santaella11}.
Individual  region  has  a  diameter of  0.35''.   We  have  retrieved
ACS-F435W, ACS-F814W, NICMOS-F110W and  NICMOS-F160W archive data from
HST.   Following  above similar  procedures,  the  optical/NIR SED  is
constructed for each region and  the stellar mass is estimated through
the stellar synthesis model \citep{Brinchmann04}.

{\noindent \bf (U)LIRG at 1$\arcsec$ resolution.}

A small sample of six local (U)LIRGs have CO J 3-2  and 880 $\mu$m images
as observed with Sub-millimeter Array (SMA) \citep{Wilson08}. We retrieved
the raw data  from the SMA archive
and re-reduced them.

For each galaxy, an aperture of about 1 kpc is  defined to enclose the whole extent
of CO emission, the integrated CO J  3-2 flux is first converted to CO
J    1-0   by    assuming   a    ratio   CO(J=3-2)/CO(J=1-0)=0.5    in the
brightness temperature \citep{Wilson08}, then  to the molecular  gas with $X_{\rm  CO}$ =
2.0x10$^{20}$ cm$^{-2}$ with an additional factor  of 1.36 to include the
Helium
\begin{equation}
M_{\rm mol} = 8.7L_{\rm CO}^{'}(3\textendash2),
\end{equation}
where  $L_{\rm CO}^{'}(3{\textendash}2$)=3.2546$\times10^{7}(\frac{\rm
S_{CO}}{\rm          Jy           km/s})(\frac{D_{\rm          L}}{\rm
Mpc})^{2}(\frac{\nu_{0}}{\rm  GHz})^{-2}(1+z)^{-1}$.   The  total  880
$\mu$m flux  within the  defined aperture is  converted to  the far-IR
luminosity by multiplying a factor of 5000 for an infrared template at
log({\rm  $L_{\rm   TIR}/L_{\odot}$})=11.7 \citep{Rieke09},  then   to  the   SFR  based   on  the
formula \citep{Kennicutt98b} with a Kroupa IMF:

\begin{equation}
 \frac{\rm SFR}{\rm M_{\odot}/yr} = 3.0\times10^{-44} \frac{L_{\rm TIR}}{\rm erg/s}.
\end{equation}

Similar to  NGC 1614, the  reduced HST images were  retrieved from
the archive  at ACS-F435W,  ACS-F814W, NICMOS-F110W,  NICMOS-F160W and
NICMOS-F222M.  Here  the NICMOS image  is registered to the  ACS image
using the field stars. No further comparison in the astrometry between
ACS and  SMA images was done,  as the accuracy ($\sim$  0.1'') is high
enough given  the adopted apertures  of 1-2$''$.  The stellar  mass is
again based on the  SED fitting \citep{Brinchmann04} to the broad SED of HST photometry. \\

\subsection{Giant Molecular Clouds In M 33}

M33 was mapped with IRAM 30 m  at CO (2-1) mainly along the major axis
with a coverage of 650 arcmin$^{2}$  at a spatial resolution of 12$\arcsec$
\citep{Gratier10}.  Circle  apertures  were defined  to  enclose
individual CO peaks,  with a diameter of 18$\arcsec$  corresponding to a
physical scale of 76 pc.  The CO  J=2-1 is converted to CO J=1-0 with
a ratio of 0.7  in the brightness temperature and to the total molecular  gas based on 
Equation~\ref{eqn_co_gas}.       The     HI      map     was      also
available \citep{Gratier10} at a resolution  of 17'', whose emission is
converted to the gas mass based on Equation~\ref{eqn_hi_gas}.  The
star formation rate  is based on the combination of  Spitzer 24 $\mu$m
image \citep{Tabatabaei07}  and   GALEX  far-UV  image   following  the
Equation~\ref{eqn_24um_fuv_sfr}, while  the stellar  mass is  based on
the     Spitzer     3.6     $\mu$m     image \citep{Dale09}     through
Equation~\ref{eqn_irac_mstar}.  Although M33 has a warped
  HI disk \citep{Rogstad76, Deul87}, all  our regions are within
  the radius where warp occurs and thus  our measurements should not
  be  affected by the warp.

\subsection{High-z Star-forming Galaxies and Starburst Galaxies}

We compiled a sample  of high-$z$ star-forming and  starburst galaxies,
and  adopted the  integrated SFRs,  gas masses  and stellar  masses as
measured  in  the  literature.   The  half-light  radius  of  the  gas
distribution or optical image is used to calculate the surface density
with the formula $\Sigma = 0.5 M/\pi/r_{\rm 0.5}^{2}$.  $X_{\rm CO J=1-0}$ =
2.0x10$^{20}$ cm$^{-2}$ are used for both star-forming and starburst galaxies.  All  star-forming
galaxies in  the literature \citep{Tacconi13} with  available half-light
radii are included  here.  The high-z starburst galaxies  are based on
several literature works as listed in  Table~\ref{tab_highz_sb}.

\section{An updated version of the extended Schmidt law} \label{sec:ES}

\begin{figure*}
\includegraphics[width=0.5\textwidth]{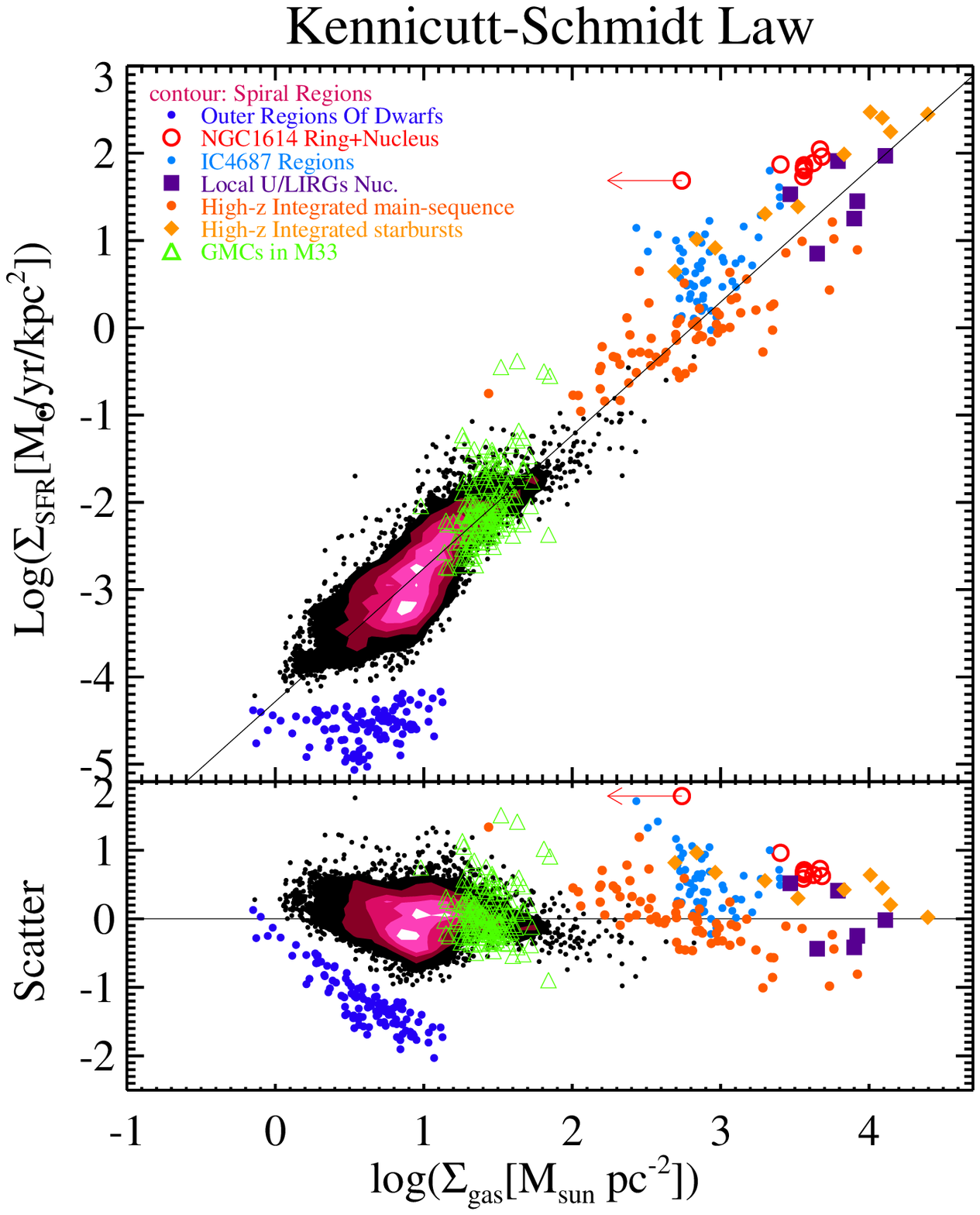}
\includegraphics[width=0.5\textwidth]{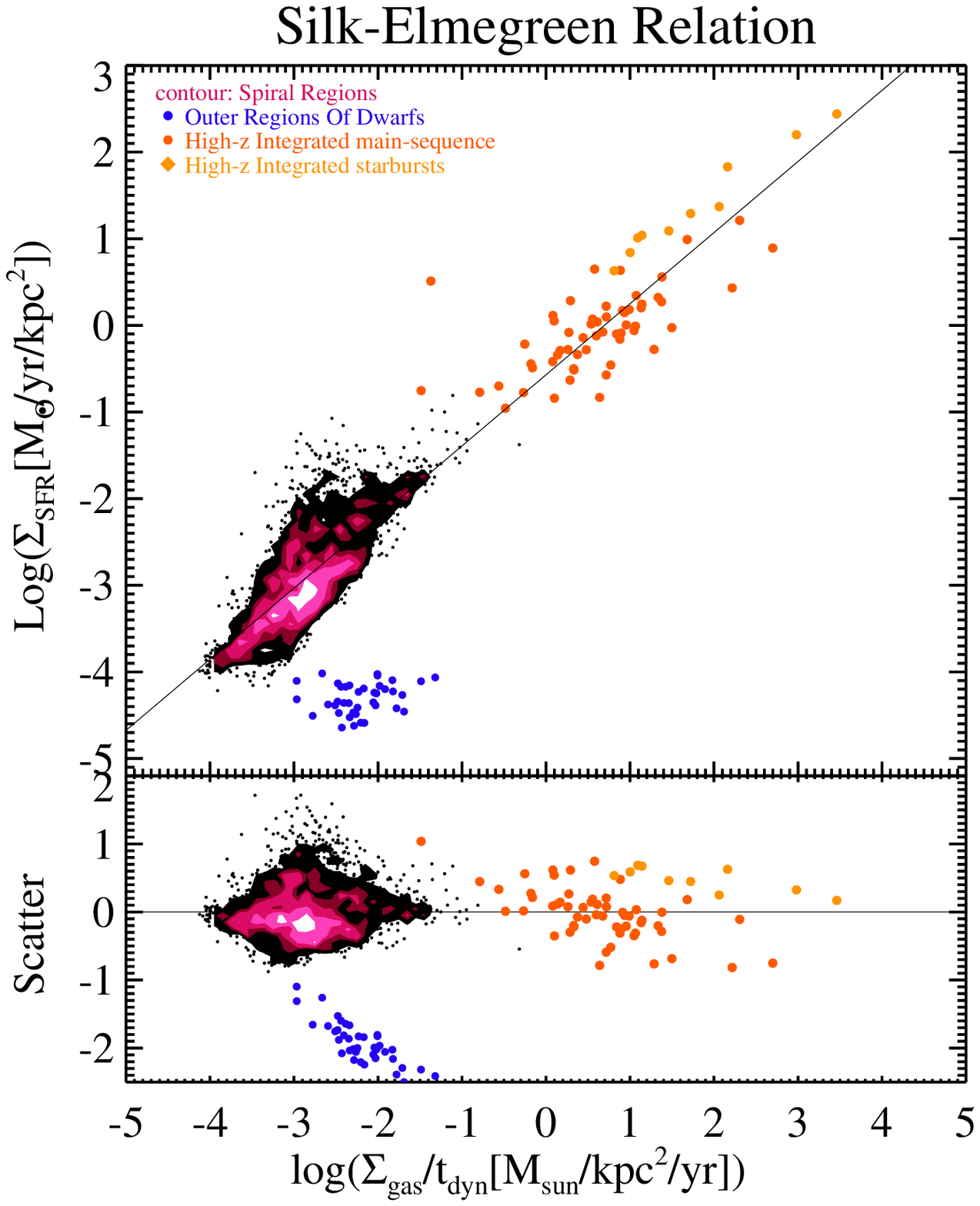}
\caption{\label{fig_KS_SE} The Kennicutt-Schmidt law (left top) and associated scatter (left bottom); the Silk-Elmegreen law (right top) and
  associated scatter (right bottom). The solid lines in all panels are the best fits. For the data of
  each sub-sample, please see the text.}
\end{figure*}

\begin{figure*}
  \centerline{ \includegraphics[width=0.8\textwidth]{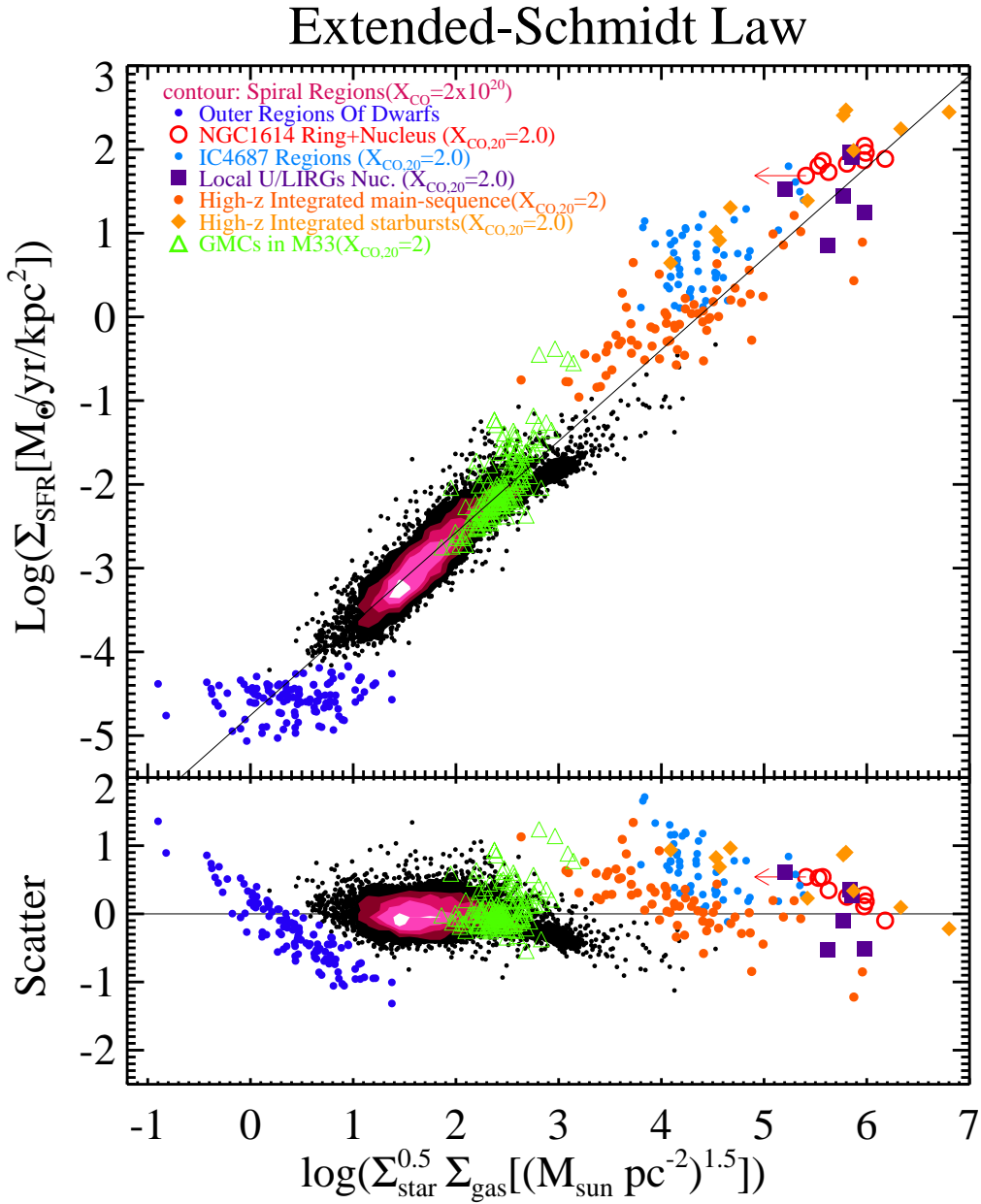} }
\caption{\label{fig_ES} The extended Schmidt law (top panel) and the associated scatter (bottom panel) where the solid line is the best
 fit. }
\end{figure*}

We present the  Kennicutt-Schmidt law, the Silk-Elmegreen  law and the
extended Schmidt  law in the  left panel of  Fig.~\ref{fig_KS_SE}, the
right   panel    of   Fig.~\ref{fig_KS_SE}    and   Fig.~\ref{fig_ES},
respectively. The data includes star  formation regions in outer disks
of dwarfs that are selected to have $\Sigma_{\rm SFR}$ below 10$^{-4}$
$M_{\odot}$/yr/kpc$^{2}$,  750 pc  grids of  spiral galaxies
from  the main  disks  to the  outer disks,  individual  GMCs in  M33,
individual star  formation regions in  LIRGs NGC  1614 and IC  4687 as
observed by ALMA, nuclear regions of  local ULIRGs as observed by SMA,
and high-z integrated main-sequence  and starburst galaxies. Only part
of the above  galaxies have rotation curves  available in the
literature to be included in the Silk-Elmegreen plot.

As found by  \citet{Shi11}, the main and outer disks  of spiral disks,
nearby (L)ULIRGs  and high-z star-forming/starburst galaxies  follow a
well defined sequence  in the extended Schmidt plane,  along with
  two additional data-sets  included in this study,  which are regions
  in the  outer disks of dwarf  galaxies and individual GMCs  in M33.
The  slope of  the  best-fitted  extended Schmidt  law  is a  slightly
super-linear:
\begin{equation}
  \Sigma_{\rm SFR} = 10^{-4.76}(\Sigma_{\rm star}^{0.5}\Sigma_{\rm gas})^{1.09},
\end{equation}
where  $\Sigma_{\rm SFR}$  is  in  M$_{\odot}$/yr/kpc$^{2}$, and  both
$\Sigma_{\rm     star}$    and     $\Sigma_{\rm    gas}$     are    in
M$_{\odot}$/pc$^{2}$.    The  1-$\sigma$   apparent  scatter   of  the
relationship  is about  0.30 dex.    Table~\ref{tab_fit_med_sigma}
lists the median offsets and standard  deviations from the best fit of
three relationships  for galaxies  of different types.  The associated
error  bars were  derived based  on the  bootstrap.  As  shown in  the
table,  the outer  disks of  dwarf  galaxies show  much larger  median
offsets in the  Kennicutt-Schmidt and Silk-Elmegreen laws  than in the
extended  Schmidt law.  Local spirals  have  no offsets  in all  three
relationships but with smallest scatters  in the extended Schmidt law.
GMCs in  M33 have  almost zero offsets  in both  Kennicutt-Schmidt and
extended Schmidt laws, but show a smaller scatter in the latter, which
is also  the case for  high-$z$ galaxies. Local (U)LIRGs  show similar
systematic  offsets   and  scatter   in  both  extended   Schmidt  and
Kennicutt-Schmidt laws.

\begin{figure}
  \centerline{ \includegraphics[width=0.5\textwidth]{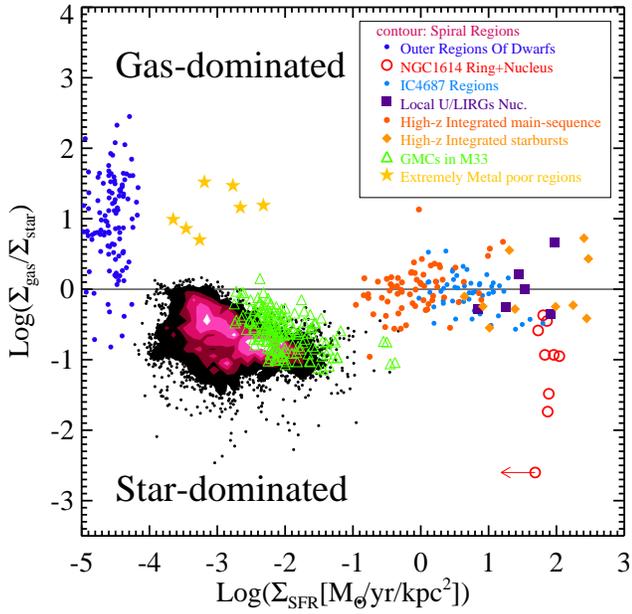} }  
\caption{\label{fig_ratio} The gas-to-stellar mass ratio as a function of the SFR surface density for our sample.}
\end{figure}

Star  formation regions  in the
outer disk of dwarf galaxies follow  the extended Schmidt law but
deviate   significantly    from   the   Kennicutt-Schmidt    law   and
Silk-Elmegreen law.  The three dwarf  galaxies presented here allow us
to  investigate extremely  low  $\Sigma_{\rm SFR}$  down to  10$^{-5}$
$M_{\odot}$/yr/kpc$^{2}$.  These regions are known to pose a challenge
not only  to the  Kennicutt-Schmidt law  \citep{Kennicutt98, Bigiel10}
but  also  to  the   Silk-Elmegreen  law  \citep{Silk97,  Elmegreen97,
  Kennicutt98,  Roychowdhury15}.  Dwarf  galaxies are  small and  thus
their orbital  timescale is much shorter  as compared to those  of LSB
galaxies  and outer  disks of  spiral  galaxies so  that they  deviate
significantly  from the  Silk-Elmegreen  relation that  holds for  LSB
galaxies and  the outer disks  of spiral galaxies.  In  addition, star
formation regions in dwarf outer  disks show high gas/star mass ratio,
ranging from 1:1  to $>$ 100:1 with  a median around 20:1  as shown in
Fig.~\ref{fig_ratio}.   The intergrated  measurements  of dwarf  outer
disks  by \citet{Elmegreen15}  confirm  the validity  of the  extended
Schmidt   law,    although   their   best-fit    slope   ($\Sigma_{\rm
  SFR}$/$\Sigma_{\rm gas}$  $\propto$ ${\sim}\Sigma_{\rm star}^{0.7}$)
is  a bit  steeper  than ours  ($\Sigma_{\rm SFR}$/$\Sigma_{\rm  gas}$
$\propto$ ${\sim}\Sigma_{\rm star}^{0.6}$).   This difference could be
due to  the different dynamic  ranges over  which the fits  were done,
with  ours  extending   into  the  starburst  regime.    As  shown  in
Fig.~\ref{fig_ES} and Table~\ref{tab_fit_med_sigma}, the  dispersion of outer disk  regions is relatively
large,  which  is  likely due  to  large  errors  (up  to 0.5  dex  at
1-$\sigma$) in  measuring the SFR,  gas mass  and stellar mass  in the
faint outer disk.

We also  investigate  the role  of the
stellar mass  at scales of giant  molecular clouds (GMCs) in  M33 with
aperture sizes  of 70  pc.  These  GMCs also
follow the power-law trend. Their  overall scatter around the best-fit
extended Schmidt  law is 0.29  dex, smaller  than 0.50 dex  around the
best-fit Kennicutt-Schmidt law.

As a summary, star formation regions in a variety of galaxy environments
from outer  disks to  Spirals and merging  galaxies, from  the sub-kpc
scale to the GMC scale, from the local to the distant Universe, all follow
a single power law among SFRs, gas masses and stellar masses. This may
indicate  the  existence  of  a  single  mechanism  that  drives  star
formation across different galactic environments.

Throughout the sample,  we adopted the same CO conversion  factor for all
sources instead of using larger  values for spirals and smaller values
for  starburst galaxies. While starburst  galaxies on  average should  have
smaller  conversion factors  than spirals, it  is still  not well
understood under  what condition a  smaller factor should be  used and
how  smaller this  factor should  be \citep[e.g.][]{Narayanan12}.  For
spatially-resolved  individual  regions  in starburst  galaxies,  this
factor  should  vary according  to  the  local condition  \citep{Xu15,
  Pereira-Santaella11}.  As a  result,  we used  the  same factor  for
simplicity.   If adopting  smaller values  for starburst  galaxies, it
will    steepen   the    slope   of    both   Kennicutt-Schmidt    law
\citep[e.g.][]{Genzel10, Liu15} and extended Schmidt law.

\section{Other outliers of Kennicutt-Schmidt law in the extended Schmidt plane}\label{sec:outlier}

Besides  the low  density  regions/galaxies,  studies have  identified
other  outliers to  the Kennicutt-Schmidt  law, suggesting  additional
parameters in regulating star formation.

\begin{figure*}
  \centerline{ \includegraphics[width=0.9\textwidth]{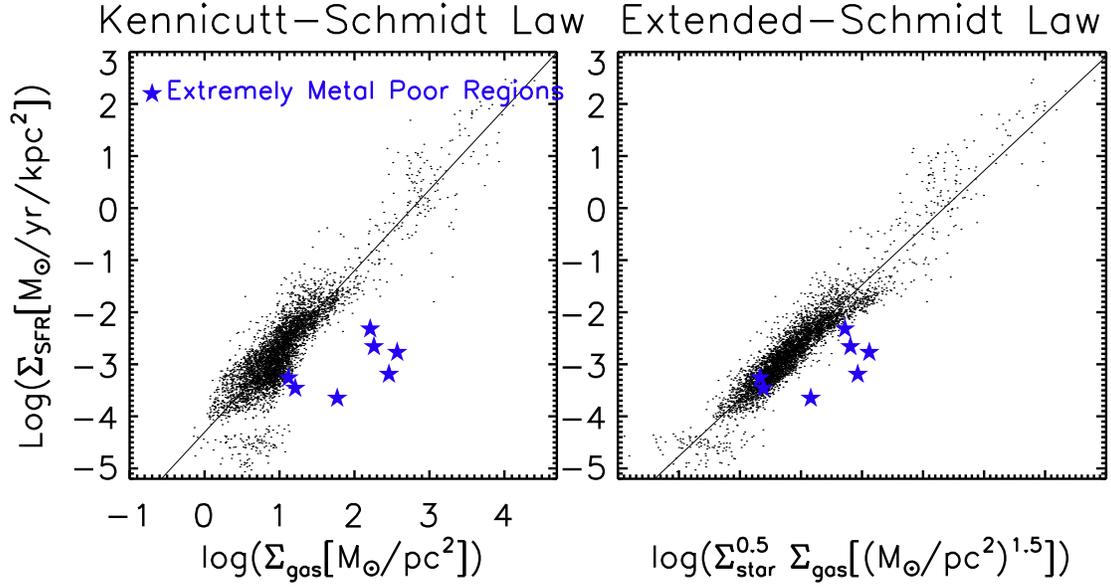} }  
\caption{\label{fig_outliers} The locations of extremely-metal poor star-formation regions on the
Kennicutt-Schmidt plane (left panel) and extended Schmidt plane (right panel). }
\end{figure*}

\citet{Shi14}  have found  that  star formation  regions of  extremely
metal-poor galaxies  with metallicities below 10\%  show significantly
lower SFR surface  densities in the Kennicutt-Schmidt  plane.   The SFR
measurements are based on the combination  of the 24 $\mu$m and far-UV
emission. Since the emission of carbon monoxide in these galaxies is extremely faint \citep{Shi15, Shi16},
the  total  cold   gas  masses  are  obtained  through
spatially-resolved dust  and HI  maps.  We  derived the  stellar mass
using the  {\it Spitzer} 3.6 $\mu$m  flux of each region  available in
\citet{Shi14}  using the  conversion by  \citet{Leroy08}. As  shown in
Fig.~\ref{fig_outliers},  the offset  of  these metal  poor regions  is
significantly  reduced,  with the  median  offset  dropping from  a
factor of 60  in the Kennicutt-Schmidt plane  to a factor of  9 in the
extended  Schmidt  plane.   \citet{Roychowdhury17} have  examined  the
integrated  measurements of  metal-poor galaxies  and found  that they
follow better the extended Schmidt law  with a mean offset of 0.01 dex
while have a mean offset of 0.3 dex in the Kennicutt-Schmidt plane.
This indicates that in metal  poor galaxies  where the
metal  plays  a  crucial  role   in  regulating star formation, the
stellar mass should be still important in regulating the star
formation processes.

\section{Theoretical Interpretations}\label{sec:model}

The validity of the extended Schmidt  law for a wide range of galactic
environments  points to  the persistent  role of  the stellar  mass in
helping to regulate star formation. Star formation can be described as
the  collapse  of a  given  gas  mass  over  a given  timescale,  i.e.
SFR=$\eta$$M_{\rm  gas}$/$t_{\rm star-formation}$,  where $\eta$  is a
constant describing  the fraction of  gas mass finally  converted into
stars.  The  extended Schmidt law  thus implies that  the star-forming
time scale is  governed by the stellar  mass, equal to the  one of the
square root of  the stellar mass surface density,  which is equivalent
to the  free fall  time for gas  in a  stellar-dominated gravitational
potential with a constant  stellar height \citep[e.g.][]{Shi11}.  Such
a  phenomenological model,  however,  does not  identify any  specific
mechanism about  the role of  the stellar  mass. Stars can  affect the
star  formation  process  through   stellar  radiation  pressure,  gas
heating,  and/or  stellar  gravity.    In  star-forming  regions,  the
radiation from old  stars (the main component of the  stellar mass) is
usually not important as compared to that from young stars.

\begin{figure}
  \centerline{ \includegraphics[width=0.45\textwidth]{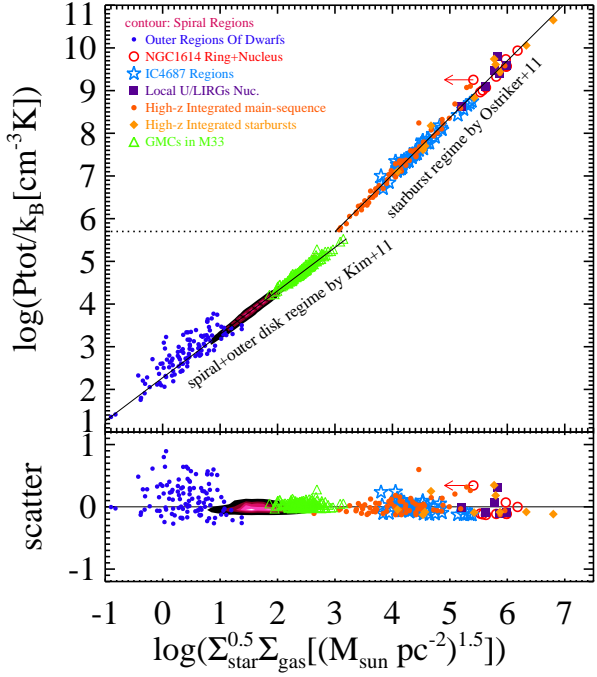} }
  \caption{\label{fig_Ptot}  The relationship of the theoretical mid-plane gas pressure with $\Sigma_{\rm star}^{0.5}\Sigma_{\rm gas}$. The two solid lines are  the best linear fits to two regimes, respectively. The horizontal dotted line
    is the division between the starburst regime and the disk regime, for each of which a different equation is used
  to estimate the mid-plane pressure (see the text).}
\end{figure}

Stellar gravity is  known to play  key roles  in converting gas
into stars. It helps to  develop the gravitational instability of cold
gas  \citep{Jog84,  Rafikov01}.  As  shown  by  \citet{Dib17}, if  the
fastest  growing  mode  of  the instability  is  associated  with  star
formation, the  observational relationship between SFR,  gas and stars
in NGC 628 could be produced  from the model.  Stellar gravity is also
important  to the  gas  pressure. As  supported  by observations,  the
pressure is related  to the fraction of gas in  the molecular phase in
which star formation takes place \citep{Elmegreen93, Blitz06}. Besides
this,  the gas  pressure also  affects the  gas thermal  and dynamical
status, making  star formation environment-dependent, a  scenario that
has been  widely explored  recently \citep{Dib11,  Monaco12, Krumholz13, Shetty13,
  Meidt16}.  Especially in  a series  of models  by \citet{Ostriker10,
  Kim11} and \citet{Ostriker11},  the stellar mass is  quantitatively incorporated
into the pressure term to predict  the SFRs.  It has been proposed that star
formation is self-regulated with  SFRs essentially proportional to the
mid-plane gas pressure that is set by the gravity of gas itself, stars
and dark matter. In spiral disks and outer disks with moderate and low
SFRs, the self-regulation is established through balancing the heating
by massive  stars with the gas  cooling that is determined  by the gas
pressure.  In  galaxies with high  SFRs, the injected momentum  to the
disk by  massive stars  is proposed  to balance the  weight of  gas.  We measured  the mid-plane
gas pressure for  our local sprials and outer disks  using the Equation 35
of \citet{Kim11}:
\begin{equation}
  \begin{split}
  P_{\rm tot, DE}  = & 1.7{\times}{10^{3}}{k_{\rm B}}f_{\rm diff}(\frac{\Sigma_{\rm gas}}{10M_{\odot}{\rm pc}^{-2}})^{2} \\
  & {\times} \{(2-f_{\rm diff})+[((2-f_{\rm diff})^{2}+37(\frac{\sigma_{\rm z, diff}}{\rm 7km/s})^{2} \\
  & {\times} (\frac{\rho_{\rm sd}}{0.1 M_{\odot}{\rm pc}^{-3}})(\frac{\Sigma_{\rm gas}}{10M_{\odot}{\rm pc}^{-2}})^{-2}]^{1/2} \}
  \end{split}           
\end{equation}
Following \citet{Kim11}, $f_{\rm diff}$ is the mass fraction of diffuse component ($<$ 50
cm$^{-3}$) and set to be 0.8 for Fig.~\ref{fig_Ptot},
$\sigma_{\rm z,  diff}$ is the  total vertical velocity  dispersion in
the  diffuse gas  and  set to  be  7.0 km/s,  $\rho_{\rm  sd}$ is  the
mid-plane density  of the stellar  disk plus  that of the  dark matter
halo, and  set to be $\Sigma_{\rm  star}$/$h$, where $h$ is  the scale
height and  set to be  300 pc for spirals and 1 kpc for the dwarf outer disks.   Plugging the $\Sigma_{\rm  star}$ and
$\Sigma_{\rm gas}$ into  the above equation, the $P_{\rm  tot, DE}$ is
plotted  against   the  term   ${\Sigma_{\rm  star}^{0.5}}{\Sigma_{\rm star}}$
in Fig.~\ref{fig_Ptot}.

For mergers and     high-$z$    galaxies     using    Equation     3    of
\citep{Ostriker11}:
\begin{equation}
  P_{\rm eff}=\frac{{\pi}G{\Sigma_{\rm gas}^{2}}}{2}(1+\chi),
\end{equation}
where    $\chi$=4$\xi_{d}\rho_{\rm   bulge}$/(3$\rho_{\rm    gas,0}$),
$\xi_{d}$  is   the  numerical   coefficient  for  the   gas  vertical
distribution but insensitive  to the exact distribution and  set to be
0.33 as suggested by \citet{Ostriker11}.   $\rho_{\rm bulge}$ is the a
mass density  of the  bulge, $\rho_{\rm gas,0}$  is the  mid-plane gas
density. We here approximate  $\rho_{\rm bulge}$/$\rho_{\rm gas,0}$ by
the observed ratio $\Sigma_{\rm star}$/$\Sigma_{\rm gas}$. This is because, for
starburst galaxies, we focused on the central regions so that $\Sigma_{\rm star}$
should represent a measurement of the stellar mass in the bulge. 

As  shown  in Figure~\ref{fig_Ptot},  in  both  regimes, the  mid-plane
pressure    is   well  approximated by the term    ${\Sigma_{\rm
    star}^{0.5}}{\Sigma_{\rm  star}}$.  This  supports  
that the stellar mass plays a  key role in regulating star formation
through its gravity.

\section{Conclusions}\label{sec:conclusion}

The extended Schmidt  law suggests a dependence of  the star formation
efficiency on the stellar mass surface density \citep{Shi11}.  In this
study we  revisit this relationship by  including spatially-resolved
measurements of star formation rates, gas masses and stellar masses of
star formation regions in a  vast range of galactic environments, from
the  outer disks  of dwarf  galaxies to  spiral disks  and to  merging
galaxies as well as individual molecular clouds in M33.  As a result of this study, we find:

(1)  Spatially-resolved star-formation  regions are  distributed in  a
tight    power-law     as    ($\Sigma_{\rm    SFR}{\propto}{\Sigma_{\rm  star}^{0.5}}{\Sigma_{\rm  gas}}$)$^{1.09}$,  which  is also  valid  for  the
integrated measurements of disk and merging galaxies at high $z$.

(2) Star formation  regions in the outer disks of  dwarf galaxies with
$\Sigma_{\rm SFR}$  down to 10$^{-5}$  M$_{\odot}$/yr/kpc$^{2}$, which
are  the outliers  of both  Kennicutt-Schmidt and  Silk-Elmegreen law,  follows
the extended Schmidt law in spite of large scatter as caused by photometric
errors.

(3)   Other   outliers  in   the   Kennicutt-Schmidt   law,  such   as
extremely-metal poor  star-formation regions, also  show significantly
reduced deviation from the extended Schmidt law.
  
(4) The mid-plane pressure is related  to ${\Sigma_{\rm  star}^{0.5}}{\Sigma_{\rm
    gas}}$,  supporting the idea that  existing stars help regulating  current  star
formation through their gravity.

\acknowledgments

We  thank   the  anonymous   referee  for  helpful   and  constructive
suggestions that improved  the quality of the paper,  Eve Ostriker for
discussions of  their model and Pereira-Santaella  for providing their
measurements of SFRs and gas mass of IC 4687. The work is supported by
the  National Key  R\&D  Program of  China  (No.  2017YFA0402704)  and
National  Natural Science  Foundation of  China (NSFC  grant 11773013,
11733002  and 11373021)  and  the Excellent  Youth  Foundation of  the
Jiangsu  Scientific Committee  (BK20150014).   K.Q.  acknowledges  the
support from NSFC (11473011 and 11590781). Z-Y.Z. acknowledges support
from ERC in  the form of the Advanced  Investigator Programme, 321302,
COSMICISM.  Based on observations obtained at the Canada-France-Hawaii
Telescope (CFHT) which is operated by the National Research Council of
Canada, the Institut National des  Sciences de l'Univers of the Centre
National de la Recherche Scientifique of France, and the University of
Hawaii.   The  GALEX  is  a  NASA Small  Explorer.   The  mission  was
developed in  cooperation with the Centre  National d’Etudes Spatiales
of France and the Korean Ministry of Science and Technology.  Based on
observations made  with the NASA/ESA Hubble  Space Telescope, obtained
from the data archive at the Space Telescope Science Institute.  STScI
is  operated  by  the  Association of  Universities  for  Research  in
Astronomy, Inc. under NASA contract NAS 5-26555.  This paper makes use
of the  following ALMA data: ADS/JAO.ALMA\#2011.0.00182.S.   ALMA is a
partnership of  ESO (representing  its member  states), NSF  (USA) and
NINS (Japan), together with NRC  (Canada), NSC and ASIAA (Taiwan), and
KASI  (Republic  of  Korea),  in  cooperation  with  the  Republic  of
Chile. The  Joint ALMA  Observatory is operated  by ESO,  AUI/NRAO and
NAOJ.  The National  Radio Astronomy Observatory is a  facility of the
National Science  Foundation operated  under cooperative  agreement by
Associated  Universities, Inc.   The  Submillimeter Array  is a  joint
project  between the  Smithsonian  Astrophysical  Observatory and  the
Academia Sinica Institute of Astronomy  and Astrophysics and is funded
by the Smithsonian Institution and the Academia Sinica.  This research
has made use  of the NASA/IPAC Extra-galactic Database  (NED) which is
operated  by the  Jet Propulsion  Laboratory, California  Institute of
Technology,  under contract  with the  National Aeronautics  and Space
Administration.

\clearpage

\begin{table}
\footnotesize
\begin{center}
\caption{\label{tab_cfht} The CFHT observations of three dwarf galaxies.}
\begin{tabular}{llllllllllllll}
\hline
Object   &  Band    &  Observing Date       & Exposure Time & seeing  \\
         &          &                       & (sec)         & ($\arcsec$) \\
\hline
Ho II    &  g-band  &  Feb. 24, 2014        & 2076         &   0.79     \\
         &  r-band  &  Feb. 24, 2014        & 2415         &   0.62     \\
Ho I     &  g-band  &  Mar. 27, 2014        & 2768         &   0.84     \\
         &  r-band  &  Mar. 28-Apr. 4, 2014 & 2760         &   0.63     \\
DDO 154  &  g-band  &  July 2, 2014         & 2768         &   0.78     \\
         &  r-band  &  July 2, 2014         & 2415         &   0.65     \\
\hline
\end{tabular}
\end{center}
\end{table}

\begin{table*}
\scriptsize
\begin{center}
\caption{\label{tab_highz_sb} The integrated properties of high-z starbursts.}
\begin{tabular}{llllllllllllll}
\hline
name            & $z$     & 2$m_{\rm a}$ & 2$m_{\rm b}$&  $\lambda_{\rm size}$  & ref$_{\rm size}$ & SFR           &  ref$_{\rm SFR}$  & $M_{\rm mol{\textendash}gas}$& ref$_{\rm gas}$ & $M_{\rm star}$ &  ref$_{\rm star}$    \\
                &         & [kpc]       &  [kpc]     &                       &                & M$_{\odot}$/yr &                  & M$_{\odot}$               &                &   $ M_{\odot}$ &                 \\
\hline                                                                
GN20            & 4.06    & 5.3         & 2.3        & 880um                 & 1     &   1860.0     &     5     & 1.3e11        &   6    &  2.3e11     &  8    \\
Aztec3          & 5.30    & 2.5         & 1.1        & 1.0mm                 & 2     &   1100.0     &     2     & 5.3e10        &   7    &  1.0e10     &  9    \\
HFLS3           & 6.34    & 2.6         & 2.4        & 1.16mm                & 3     &   2900.0     &     3     & 1.0e11        &   3    &  3.7e10     &  3    \\
SMMJ123549+6215 & 2.20    & 1.8         & 1.8        & CO                    & 4     &   897.0      &     4     & 7.1e10        &   4    &  1.2e11     &  4    \\
SMMJ123634+6212 & 1.22    & 8.2         & 8.2        & CO                    & 4     &   465.0      &     4     & 5.2e10        &   4    &  6.6e10     &  10   \\
SMMJ123707+6214 & 2.49    & 5.6         & 5.6        & CO                    & 4     &   508.0      &     4     & 3.4e10        &   4    &  1.2e11     &  4     \\
SMMJ131201+4242 & 3.41    & 6.0         & 6.0        & CO                    & 4     &   465.0      &     4     & 5.2e10        &   4    &  9.1e10     &  11    \\
SMMJ131232+4239 & 2.33    & 4.0         & 4.0        & CO                    & 4     &   508.0      &     4     & 5.0e10        &   4    &  1.4e10     &  11    \\
SMMJ163650+4057 & 2.39    & 4.8         & 4.8        & CO                    & 4     &   886.0      &     4     & 1.2e11        &   4    &  2.3e11     &  4     \\
SMMJ163658+4105 & 2.45    & 1.6         & 1.6        & CO                    & 4     &   1124.0     &     4     & 1.0e11        &   4    &  2.6e11     &  4     \\
\hline
\end{tabular}
\tablecomments{1-\cite{Hodge15}; 2-\cite{Riechers14}; 3-\cite{Riechers13}; 4-\cite{Genzel10}; 5-\cite{Hodge15};
  6-\cite{Carilli11}; 7-\cite{Riechers10}; 8-\cite{Daddi09}; 9-\cite{Capak11}; 10-\cite{Hainline11}; 11-\cite{Hainline11}}
\end{center}
\end{table*}

\begin{table*}
\footnotesize
\begin{center}
\caption{\label{tab_fit_med_sigma} The median offsets and standard deviations of galaxies from the extended Schmidt, the Kennicutt-Schmidt, and Silk-Elmegreen laws.}
\begin{tabular}{llllllllllllll}
\hline
Galaxy Types               &  Extended Schmidt Law    &  Kennicutt-Schmidt Law  & Silk-Elmegreen Law      \\
                           &  (median, $\sigma$)    &   (median, $\sigma$)      &      (median, $\sigma$)          \\
\hline
outer disks of dwarfs      &  -0.26$\pm$0.04,  0.48$\pm$0.03      &   -1.20$\pm$0.04,   0.43$\pm$0.03    & -1.92$\pm$0.05, 0.31$\pm$0.04         \\         
local spirals              &  -0.01$\pm$0.00,  0.24$\pm$0.003     &    0.02$\pm$0.01,   0.33$\pm$0.004   &  0.03$\pm$0.01, 0.39$\pm$0.01        \\     
GMCs in M33                &   0.06$\pm$0.02,  0.29$\pm$0.02      &    0.01$\pm$0.04,   0.50$\pm$0.06    &                                      \\             
local (U)LIRGs             &   0.56$\pm$0.06,  0.46$\pm$0.04      &    0.53$\pm$0.05,   0.44$\pm$0.04    &                                      \\     
high-z MS and SB galaxies  &   0.06$\pm$0.02,  0.29$\pm$0.02      &    0.08$\pm$0.05,   0.46$\pm$0.04    &  0.08$\pm$0.06, 0.47$\pm$0.07         \\     
\hline
\end{tabular}
\tablecomments{The uncertainties of the median offsets and standard deviations are obtained by the bootstrap.}
\end{center}
\end{table*}

\clearpage

\end{document}